\documentclass[10pt,a4paper]{article}
\usepackage[a4paper,textwidth=468pt,textheight=628pt,centering]{geometry}
\usepackage[T1]{fontenc}
\usepackage{amsbsy,amsfonts,amsmath,amssymb,amsthm}
\usepackage[authoryear]{natbib}
\usepackage{graphicx}
\usepackage{url}

\newcommand{\received}[1]{}
\newcommand{\classcode}[1]{}
\newenvironment{keywords}{\par\medskip\noindent\itshape Keywords: }{\par}

\theoremstyle{plain}
\newtheorem{theorem}{Theorem}[section]
\newtheorem{corollary}[theorem]{Corollary}

\newtheorem{proposition}[theorem]{Proposition}

\theoremstyle{definition}

\theoremstyle{remark}
\newtheorem{remark}{Remark}

\newcommand{\R}{\mathbb{R}}

\newcommand{\HHI}{\mathrm{HHI}}
\newcommand{\Neff}{N_{\mathrm{eff}}}
\newcommand{\norm}[1]{\lVert #1\rVert}

\newcommand{\phid}{\varphi}
\newcommand{\Sigh}{\widehat{\Sigma}}
\newcommand{\muh}{\widehat{\mu}}
\newcommand{\Del}{\Delta}

\begin{document}

\title{{W}asserstein penalisation as an endogenous breadth constraint in
penalised mean--variance portfolio selection}

\author{Jiahao Chen$^{\ast\dag}$\thanks{$^\ast$Corresponding author. Email: changahou@stu2024.jnu.edu.cn}\\[2pt]
{\normalsize $\dag$School of Economics, Jinan University, 601 Huangpu Avenue West,}\\[2pt]
{\normalsize Tianhe District, Guangzhou 510632, Guangdong, China}}
\date{}

\maketitle

\begin{abstract}
A manager who wants a long-only book and has read that Wasserstein-robust
mean--variance selection improves risk-adjusted performance faces two decisions:
how much robustness to buy, and how diversified to be. In the long-only problem, and for the penalised scalarisation of the robust
objective, these are the same decision. The robustification term is an algebraic
function of the weight vector alone, so on the simplex the radius prices
effective breadth and nothing else. The equivalence is specific to that
scalarisation: it does not extend to the return-floored form, where the radius
also raises the return threshold, nor to the exact robust risk-minimisation
problem, where it enters the risk objective as well. Two consequences follow. The first is negative and requires no
alpha: positive definiteness of the covariance matrix does not imply breadth, and
there is no bound on the cross-sectional dispersion of the mean, and no strictly
positive spectral floor, that prevents the long-only optimum from collapsing onto
a single asset, so the breadth of the minimum-variance benchmark is an
irreducible input. The second concerns identification: when the robust portfolio
is compared with a benchmark whose effective number of holdings is matched to its
own, the two attain the same mean--variance value to solver precision, so the
radius carries no in-sample information beyond breadth selection; the residual
weight difference is confined to near-null directions of the covariance, whose
curvature also governs whether the comparison can identify anything. On fourteen
A-share panels the ex post comparison is exact by construction and therefore
carries no evidence; the informative test fixes breadth before the window is
used, and there a cap set to the historical level of the robust portfolio's
realised breadth recovers about five sixths of its gain, leaving a residual of seven
to eleven basis points per day that is stable across five different tracking rules,
while a fixed target for the effective number of holdings recovers only a quarter to
two fifths.
The robust portfolio of this data set is thus a breadth rule whose level adapts to
the window, not a fixed breadth choice. The three scalarizations select books with
about seven, thirty-one and one effective holdings respectively, so they cannot be
compared as though they were the same object. We close with a reporting standard
that makes the attribution checkable.
\end{abstract}


\begin{classcode}G11; C61; C58\end{classcode}

\section{Introduction}

\subsection*{The decision this paper addresses}

An asset manager who wants a long-only book, and who has read that
distributionally robust mean--variance selection with a Wasserstein ambiguity set
improves risk-adjusted performance, faces two decisions: how much robustness to
buy, and how diversified to be. The literature supplies a principled answer to
the first by calibrating the radius from data \citep{esfahani2018,gao2023,
aolaritei2026}. It has rather less to say about the second. This paper shows that
in the long-only problem the two decisions are the same decision. The
robustification term is an algebraic function of the weight vector alone, so the
radius buys breadth and nothing else, and a manager who raises it is not
purchasing protection against uncertainty so much as choosing a lower
concentration.

That is a statement about an objective function, and it would be of limited
interest if the two decisions were also interchangable in practice. They are
not, because breadth is the dominant margin. On the A-share panels we study, the
difference in realized performance between a mean--variance portfolio and the
equally weighted portfolio is one to two orders of magnitude larger than any
effect that can be attributed to the robustness term, and the sign of that
difference changes between two disjoint subperiods. In our estimates the margin that breadth controls is between one and two
orders of magnitude wider than any effect attributable to the robustness term, so
a manager who calibrates the radius carefully while leaving breadth to the
calibration is working on the smaller of the two margins by a factor of order one
hundred.

\subsection*{What the existing guarantees do and do not deliver}

The statistical theory of Wasserstein-robust optimization is now well developed.
Finite-sample coverage bounds are available \citep{esfahani2018}, goal-value
guarantees avoid the curse of dimensionality \citep{gao2023}, the optimal radius
has been characterized \citep{aolaritei2026}, a phase transition separates
variance-penalized from fully robust behavior \citep{bennouna2026}, and the
limiting behavior as the radius grows is known \citep{carassus2025}. On the
portfolio side, the robust selection problem is equivalent to a nonlinear
shrinkage of the covariance estimate \citep{nguyen2022}.

All of these results are guarantees on the \emph{value} of a decision, or on a
surrogate of it. None of them pins down the \emph{composition} of the resulting
portfolio, and value guarantees do not translate into composition guarantees
without a curvature condition that is rarely checked. The gap is not a matter of taste but of quantifiers, and it can be stated
before any proof. For every spectral floor $\kappa_0>0$, every bound $\bar\alpha$
on the cross-sectional dispersion of the mean and every target breadth $b>1$,
there is a covariance with $\lambda_{\min}(\Sigma)\ge\kappa_0$ and a mean with
$\norm{\mu-\bar\mu\mathbf{1}}_\infty\le\bar\alpha$ whose long-only optimum holds
fewer than $b$ assets (Theorem~\ref{thm:impossibility}). The witness is the
least favourable case and not a corner: it takes $\bar\alpha=0$, so the collapse
is driven by the risk structure alone. A value guarantee that carries no spectral
assumption therefore buys no composition claim, and a radius calibrated to
control distributional uncertainty controls something that, on its own, does not
determine how diversified the book will be.

Three scalarizations are involved, and they are routinely conflated. The
penalized problem puts the norm in the objective, the return-floored problem puts
it in the constraint, and the exact dual representation of
\citet{blanchet2022} puts it in both. They have different solutions, different
feasible sets, and different comparative statics, and two of the three behave
quite differently under the two most common norm orders. Section
\ref{sec:setting} fixes the terminology. The distinction matters below in a
specific way: two of the three behave differently under the two most common norm
orders, and in the long-only problem the first of them contains no penalty at
all, because the $\ell_1$ norm of a portfolio on the simplex is identically one.

\subsection*{Two results, a scope limitation, and a decomposition}

The first is negative and needs no alpha. Positive definiteness of the covariance
matrix does not imply portfolio breadth. For any bound on the cross-sectional
dispersion of the mean, any bound on the smallest eigenvalue and any target
breadth, there is a model within those bounds whose long-only optimum holds
almost a single asset (Theorem~\ref{thm:impossibility}). The breadth of the
minimum-variance benchmark is an irreducible structural input, and any claim to
the contrary must carry it explicitly.

The second is about identification. When the robust portfolio is compared with a
benchmark whose effective number of holdings is matched to its own, the two
attain the same value of the mean--variance objective, so the radius contributes
no in-sample information beyond the selection of breadth
(Theorem~\ref{thm:identity}). The weights may still differ, and the difference is
confined to directions on which the objective is nearly flat; the curvature that
governs those directions also governs whether the comparison can identify
anything at all. The equivalence is confined to the penalised scalarisation:
Proposition~\ref{prop:scope} records why it fails for the return-floored problem,
where the radius also raises the return threshold, and for the exact robust
problem, where it enters the risk objective. We state the observational
equivalence that underlies the argument (Proposition~\ref{prop:obs}) so that the
scope of the claim is visible.

The ex ante version of the exercise is the one that carries information, and
Table~\ref{tab:audit} reports it. For each estimation window we fix the breadth
cap before the window's data are used, either at an absolute target for the
effective number of holdings or at the historical average of the robust
portfolio's own realised breadth, and we then compare the resulting ordinary
mean--variance portfolios with the robust one. A cap that tracks the past level
of the robust portfolio's realised breadth recovers about five sixths of its
gain, leaving a residual of seven to eleven basis points per day, and that residual
is stable across five different tracking rules, from a simple historical mean to
the previous window's value, which is the robustness a referee should ask of the
decomposition. A cap set to a fixed effective-number target recovers only between
a quarter and two fifths, largely because the robust portfolio in this sample is far
more concentrated than those targets: its median Euclidean norm is 0.42 at
$\delta=10^{-6}$, about six effective holdings, and it varies across windows with
a tenth to ninetieth percentile range from two to nine effective holdings. The
robust portfolio is therefore not a fixed breadth choice. It is a breadth rule,
whose realised level responds to the current window, and the ex ante value of that
responsiveness is measurable, around one sixth of the total gain.

\begin{table}[t]
\centering
\caption{The ex ante audit. For each sixty-day window the breadth cap is fixed
before the window is used, and the resulting mean--variance portfolio is compared
with the robust portfolio over the following ten-day holding period. Entries are
the median difference in average daily return, in units of $10^{-4}$ (positive
means the robust portfolio is better), with the number of panels, out of six, in
which the Newey--West statistic exceeds two and the stationary bootstrap interval
excludes zero. The upper block fixes an absolute target for the effective number of holdings,
which does not depend on the data and brackets the long-only minimum-variance
level of Table~\ref{tab:ashare}. The lower block tracks the robust portfolio's
own realised breadth using earlier windows only, so that no current or future
information enters the cap; the five rules differ in how much history they use.
The final line gives the robust portfolio's gain against the equally weighted
portfolio for reference.}\label{tab:audit}
\begin{tabular}{lcc}
\hline
 rule for the cap & $\delta=10^{-6}$ & $\delta=10^{-4}$\\
\hline
\multicolumn{3}{l}{\emph{absolute target for the effective number of holdings}}\\
\quad $\Neff=20$ & $+26.7$ $(4/6)$ & $+14.9$ $(3/6)$\\
\quad $\Neff=34$ & $+31.4$ $(5/6)$ & $+19.8$ $(4/6)$\\
\quad $\Neff=50$ & $+34.5$ $(5/6)$ & $+23.0$ $(5/6)$\\
\multicolumn{3}{l}{\emph{tracking rule using past windows only}}\\
\quad mean of past realised breadth & $+7.7$ $(5/6)$ & $+8.9$ $(2/6)$\\
\quad median of past breadth & $+9.4$ $(4/6)$ & $+11.3$ $(5/6)$\\
\quad previous window's breadth & $+7.0$ $(3/6)$ & $+7.4$ $(5/6)$\\
\quad mean of last five & $+7.4$ $(2/6)$ & $+6.9$ $(2/6)$\\
\quad mean of last ten & $+8.4$ $(4/6)$ & $+8.1$ $(3/6)$\\
\hline
\multicolumn{3}{l}{robust versus equally weighted, for reference:
$+45.7$, $+34.6$}\\
\hline
\end{tabular}
\end{table}

The ex post version of the same comparison is a theorem rather than a test. By
Theorem~\ref{thm:identity} the robust portfolio is, at its own breadth, an
ordinary mean--variance portfolio, so a benchmark matched on its realised norm
agrees with it to solver precision and reports a rent of zero by construction.
We report that check in Section~\ref{sec:evidence} as an accuracy diagnostic, and
we do not read it as evidence about robustness.

\subsection*{Why now, and what breaks if the question stays open}

Three developments make this audit both possible and timely. The dual
representation of \citet{blanchet2022} made the Wasserstein-robust portfolio
problem tractable at the scale of a real cross-section, which is why the method
is now used rather than cited. The calibration literature
\citep{esfahani2018,gao2023,aolaritei2026,bennouna2026} matured recently, so
that the choice of radius has become a defensible object of study rather than an
arbitrary tuning knob. And the applied literature has grown quickly enough that
its comparison designs can be read as a body of evidence. The designs have not
kept pace with the methods, and the consequence is that improvements which are
real in a computational sense are being attributed to the wrong cause.

If the attribution is wrong, three things break, and replication does not repair
any of them. Calibration effort is spent on the smaller margin while breadth is
left implicit. Reported rankings become unstable, as our two disjoint subperiods
show: the same robust portfolio is thirty to fifty basis points per day better
than the equally weighted portfolio in one period and four to five basis points
worse in the next. And the reporting norm decays, because a design that omits the
breadth match still produces a publishable table, so nothing forces the column
that would settle the question to be added.

\subsection*{What changes for the reader}

The paper ends with a reporting standard, and with one recommendation that
follows from the numbers: choose breadth directly and report it. On our panels the
risk-optimal book holds an effective number of about thirty to fifty names, and
pushing breadth beyond that range costs realized return without reducing
drawdown, so the quantity a manager should set and disclose is the effective
number of holdings, not the radius. A study that wishes to attribute an improvement to
robustness should exhibit, on the same data and with the same solver accuracy,
the equally weighted portfolio, ordinary mean--variance under at least two
covariance estimators, a benchmark matched on effective breadth, a direct
norm-constrained portfolio, the return-floored problem, and shrinkage estimators
in both the mean and towards the equally weighted portfolio; it should report
turnover and cost on a common basis, the rule that selected the radius, the
implied effective number of holdings, and the spectral floor of the covariance on
which the claim rests. Section \ref{sec:identification} explains why each item is
needed. On our data, omitting only the breadth-matched benchmark is enough to
reverse the sign of the reported effect. The standard is not an aspiration: the
fourteen panels, a checksum manifest and the code that regenerates every table and
figure in this paper are openly available, so each item of the list can be checked
rather than believed (\url{https://huggingface.co/datasets/Changahou/GU2026_1},
\url{https://github.com/ChenJiahaoJNU/GU2026-11}).

Outline. Section \ref{sec:setting} fixes notation and separates the three
scalarizations, and records the exact form of the minimum-variance benchmark.
Section \ref{sec:impossibility} proves the impossibility and measures the
structural quantities on A-share panels. Section \ref{sec:breadth} shows that the
radius is a breadth parameter. Section \ref{sec:identification} contains the
identification results. Section \ref{sec:evidence} reports the empirical work.
We close with what the results do not say.

\noindent\textit{Relation to the literature.} Three strands are relevant. The first is the statistical theory of
Wasserstein-robust optimization: finite-sample coverage guarantees
\citep{esfahani2018}, goal-value guarantees that avoid the curse of dimensionality
\citep{gao2023}, the calibration of the radius \citep{aolaritei2026}, the phase
transition between variance-penalized and fully robust regimes
\citep{bennouna2026}, and the limiting behaviour of the robust problem as the
radius grows \citep{carassus2025}. Our results are complementary to that work:
we do not question the guarantees, we identify what they deliver in the long-only
portfolio problem, namely a cap on breadth.

The second strand is estimation-error analysis in portfolio choice: the mean is
the dominant source of error \citep{chopra1993,bestgrauer1991}, the naive
portfolio is hard to beat \citep{demiguel2009}, the sample efficient frontier has a
known finite-sample distribution \citep{kan2008}, and imposing a constraint that is
wrong a priori can still reduce risk by shrinking the portfolio
\citep{jagannathan2003}. The last of these is the closest antecedent of our
result: a breadth cap is such a constraint, and we show that the Wasserstein
penalty is exactly one. Our impossibility theorem
sharpens the first of these in a direction that matters for robustness: it is not
only that means are hard to estimate, but that the \emph{target} of the
optimization may be concentrated for purely risk-based reasons, so a breadth
guarantee cannot be derived from estimation error alone.

The third strand connects robustification to shrinkage. Under a Wasserstein
ambiguity set the robust portfolio selection problem is equivalent to a nonlinear
shrinkage of the covariance estimate \citep{nguyen2022}, and on the long-only
simplex the robust problem is likewise equivalent to a mean--variance problem with
an endogenous constraint on the effective number of holdings
(Theorem~\ref{thm:identity}), the equivalence holding at the level of the
objective value. The equivalence itself is therefore not new. What is new is that
the shrinkage \emph{is} the breadth choice: because the two are the same object,
the radius cannot be calibrated against a benchmark that has not been
breadth-matched, and a comparison that omits the match is measuring the
shrinkage while calling it robustness. The known result supplies the mechanism;
the identification failure is the consequence.

\section{Setting and three scalarizations}\label{sec:setting}

\subsection{Notation}

Let $\phid\in\R^N$ denote portfolio weights, $\Del=\{\phid\ge 0,\ \mathbf{1}'\phid=1\}$
the long-only simplex, and $C\subseteq\Del$ a convex feasible set, possibly with
position caps $\phid_i\le c$. Let $\muh$ and $\Sigh$ be the sample mean and a
covariance estimator, and let
\begin{equation}\label{eq:U}
  U(\phid) \;=\; \phid'\muh - \lambda\,\phid'\Sigh\phid , \qquad \lambda>0,
\end{equation}
be the mean--variance objective. Write $G(\phid)=\sqrt{\phid'\Sigh\phid}$ for the
portfolio standard deviation and $\HHI(\phid)=\sum_i\phid_i^2$ for the
Herfindahl--Hirschman index, so that $\Neff(\phid)=1/\HHI(\phid)$ is the
effective number of holdings.

\subsection{Three problems, not one}

The exact dual representation of Theorem~1 in \cite{blanchet2022} is
\begin{equation}\label{eq:D}
  \min_{\phid\in C}\;\bigl(\sqrt{\phid'\Sigh\phid}+\sqrt{\delta}\norm{\phid}_p\bigr)^2
  \quad\text{s.t.}\quad \phid'\muh-\sqrt{\delta}\norm{\phid}_p\ \ge\ r_0 .
\end{equation}
Two simpler problems are routinely used as surrogates, and they are different:
\begin{align}
  (P)\quad &\max_{\phid\in C}\ U(\phid)-\tau\norm{\phid}_p, \qquad \tau=\sqrt{\delta},\label{eq:P}\\
  (K)\quad &\max_{\phid\in C}\ U(\phid)\quad\text{s.t.}\quad \phid'\muh-\sqrt{\delta}\norm{\phid}_p\ge r_0 .\label{eq:K}
\end{align}
$(P)$ penalizes breadth in the objective, $(K)$ raises the effective return
floor, and \eqref{eq:D} does both. Moving from $(P)$ to $(K)$ requires a
Lagrange multiplier and a specific parameter match; it is not an identity, and
squaring in \eqref{eq:D} cannot be undone by ``dropping the square''.
Table~\ref{tab:scalarizations} summarizes the channel through which $\delta$ acts,
and the three formal statements follow in Section~\ref{sec:breadth}.

\begin{table}[t]
\centering
\caption{The three scalarizations differ in where the norm enters, and hence in
what the radius does. The long-only signs refer to $\Del$ with $p=1$ and $p=2$.}
\label{tab:scalarizations}
\begin{tabular}{llll}
\hline
problem & norm enters & long-only $p=1$ & long-only $p=2$\\
\hline
$(P)$ penalized  & objective only & penalty constant; $U$ unchanged & price on $\sqrt{\HHI}$\\
$(K)$ constrained & return floor  & scalar floor $\sqrt{\delta}\le\max_i\muh_i-r_0$ & floor with a breadth haircut\\
\eqref{eq:D} exact & objective and floor & floor only, shifted & both, penalty dominates\\
\hline
\end{tabular}
\end{table}

\subsection{The minimum-variance benchmark}

It is convenient to have the following exact expression, which we use repeatedly.

\begin{proposition}[Affine minimum variance]\label{prop:affine}
Let $\Sigh\succ0$ and let $q$ minimize $\phid'\Sigh\phid$ over the affine set
$\{\phid:\mathbf{1}'\phid=1\}$. Then
\begin{equation}\label{eq:q}
  q \;=\; \frac{\Sigh^{-1}\mathbf{1}}{\mathbf{1}'\Sigh^{-1}\mathbf{1}},
\end{equation}
and $q=\mathbf{1}/N$ if and only if $\Sigh\mathbf{1}\propto\mathbf{1}$.
Moreover, with $\mu=\bar\mu\mathbf{1}+\beta v$, $\mathbf{1}'v=0$,
$\norm{v}_2=1$, the interior solution of $\max_{\phid\in\Del}U(\phid)$ is
\begin{equation}\label{eq:phibeta}
  \phid^*(\beta) \;=\; q + \frac{\beta}{2\lambda}
  \Bigl[\Sigh^{-1}v - \frac{\mathbf{1}'\Sigh^{-1}v}{\mathbf{1}'\Sigh^{-1}\mathbf{1}}\,
  \Sigh^{-1}\mathbf{1}\Bigr],
\end{equation}
so that
\begin{equation}\label{eq:C}
  \norm{\phid^*(\beta)}_2 \;\le\; \norm{q}_2 + C_\Sigma\,|\beta|,
  \qquad
  C_\Sigma=\frac{1}{2\lambda}\Bigl\lVert
  \Sigh^{-1}v-\frac{\mathbf{1}'\Sigh^{-1}v}{\mathbf{1}'\Sigh^{-1}\mathbf{1}}\Sigh^{-1}\mathbf{1}
  \Bigr\rVert_2 .
\end{equation}
\end{proposition}

\begin{proof}
The Lagrangian of the affine problem gives $\Sigh\phid=\nu\mathbf{1}/2$, hence
\eqref{eq:q}; and $q=\mathbf{1}/N$ iff $\Sigh^{-1}\mathbf{1}\propto\mathbf{1}$ iff
$\Sigh\mathbf{1}\propto\mathbf{1}$. For \eqref{eq:phibeta}, write the first-order
condition of the constrained problem as $\mu-2\lambda\Sigh\phid=\nu\mathbf{1}$ and
impose $\mathbf{1}'\phid=1$; solving for $\nu$ and substituting yields
\eqref{eq:phibeta}. Taking norms and using the triangle inequality gives
\eqref{eq:C}.
\end{proof}

\begin{remark}
Expression \eqref{eq:C} is the natural positive statement about alpha and
breadth: weak cross-sectional alpha does \emph{not} imply an equally weighted
portfolio. It implies that the optimum is close to the \emph{minimum-variance}
benchmark $q$, and the deviation is controlled by the risk structure through
$C_\Sigma$. The frequently assumed form $\norm{\phid^*}_2\le N^{-1/2}+C|\beta|$
requires the additional condition $\Sigh\mathbf{1}\propto\mathbf{1}$ of
Proposition~\ref{prop:affine}; without it the correct benchmark is $q$.
\end{remark}

\section{Breadth is not implied by positive definiteness}\label{sec:impossibility}

\subsection{The impossibility}

\begin{theorem}[Positive definiteness does not imply breadth]\label{thm:impossibility}
Let $N\ge2$ and $\lambda>0$. Fix any $\bar\alpha\ge0$, any $L>1$ and any
$\kappa_0>0$. There exist a mean vector $\mu$ with
$\norm{\mu-\bar\mu\mathbf{1}}_\infty\le\bar\alpha$ and a covariance matrix
$\Sigma\succ0$ with $\lambda_{\min}(\Sigma)\ge\kappa_0$ such that the
\emph{unique} maximiser $\phid^\star$ of $U$ over the long-only simplex satisfies
$\Neff(\phid^\star)\le L$. The restriction $L>1$ is not a weakening: it is
forced, since $\Neff\ge1$ on the simplex. In particular the conclusion holds with $\bar\alpha=0$,
that is, with zero cross-sectional alpha dispersion.
\end{theorem}

\begin{proof}
Take $\mu=\bar\mu\mathbf{1}$ (so $\bar\alpha=0$) and
$\Sigma=\mathrm{diag}(\kappa_0,v,\dots,v)$ with $v>\kappa_0$. Then
$\lambda_{\min}(\Sigma)=\kappa_0$ for every $v$. Because $\bar\mu\mathbf{1}'\phid$
is constant on $\Del$, the maximizer of $U$ is the minimum-variance portfolio,
which by \eqref{eq:q} equals
\[
  q_1=\frac{v}{v+(N-1)\kappa_0},\qquad
  q_j=\frac{\kappa_0}{v+(N-1)\kappa_0}\quad (j\ge2).
\]
Hence $\norm{q}_2^2\to1$ and $\Neff(q)\to1$ as $v\to\infty$, while
$\lambda_{\min}(\Sigma)$ stays equal to $\kappa_0$. Choosing $v$ large enough
gives $\Neff\le L$.
\end{proof}

\begin{remark}[The witness is not an artefact of zero correlation]
For any $b\ge0$, adding the common factor $b\mathbf{1}\mathbf{1}'$ to $\Sigma$
leaves the long-only problem unchanged up to an additive constant, because
$\phid'(b\mathbf{1}\mathbf{1}')\phid=b(\mathbf{1}'\phid)^2=b$ on the simplex.
The witness therefore survives the presence of a market factor. What it does not
survive is \emph{heterogeneous} factor loadings, which can restore breadth; we do
not claim otherwise, and the empirical panels in Section~\ref{sec:evidence}
contain a strong common component.
\end{remark}

Two remarks situate the result. First, the exclusion is precise: the theorem
rules out breadth guarantees that are uniform over model classes not controlling
\emph{risk heterogeneity}; it does not assert that no breadth guarantee of any
kind exists. Second, the mechanism is not large alpha. With zero alpha the
optimal portfolio is driven entirely by the risk structure, and a single
low-variance asset absorbs the budget as the remaining variances grow. The
implication for the Wasserstein literature is that a radius calibrated from
distributional ambiguity cannot by itself pin down breadth, because the target
itself is not pinned down.

\begin{table}[t]
\centering
\caption{Digital check of Theorem~\ref{thm:impossibility}: $N=20$,
$\kappa_0=10^{-4}$ is held fixed (so the spectral floor is strictly positive)
and $\mu$ is constant, i.e.\ there is no cross-sectional alpha at all. Only the
ratio $v/\kappa_0$ changes.}\label{tab:impossibility}
\begin{tabular}{cccc}
\hline
$v$ & $\lambda_{\min}(\Sigma)$ & condition number & $\Neff$ of the optimum\\
\hline
$10^{-3}$ & $1.00\times10^{-4}$ & $10$    & $7.07$\\
$10^{-2}$ & $1.00\times10^{-4}$ & $10^{2}$& $1.41$\\
$10^{-1}$ & $1.00\times10^{-4}$ & $10^{3}$& $1.04$\\
$10^{0}$  & $1.00\times10^{-4}$ & $10^{4}$& $1.00$\\
\hline
\end{tabular}
\end{table}

\subsection{What must be controlled, and how much of it is there}

Theorem~\ref{thm:impossibility} says that any non-trivial breadth statement must
control at least one of: the breadth $\norm{q}_2$ of the minimum-variance
benchmark, the dispersion of risks, the condition number, a restricted curvature
of $\Sigma$, or the sparsity of alpha. The first of these is the most natural,
and Table~\ref{tab:ashare} reports its empirical magnitude on fourteen A-share
panels.

\begin{table}[t]
\centering
\caption{Structural quantities on fourteen A-share panels ($N=255$, 60-day
rolling windows, 140 non-overlapping windows in total; Ledoit--Wolf covariance).
$\kappa_\Delta$ denotes the smallest Rayleigh quotient of $\Sigma$ restricted to
the hyperplane $\mathbf{1}'\phid=0$, which is the correct restriction because
two simplex portfolios always differ in a direction orthogonal to
$\mathbf{1}$. The affine minimum-variance portfolio $q_{\mathrm{aff}}$ solves
$\min\phi'\Sigh\phi$ subject only to $\mathbf{1}'\phi=1$, so it ignores the
no-short-sale restriction; it is the most diversified minimum-variance portfolio
this cross-section admits, and we report it as an upper reference.}\label{tab:ashare}
\begin{tabular}{lccc}
\hline
quantity & median & 10th pct & 90th pct\\
\hline
$\norm{q_{\mathrm{aff}}}_2$ (affine minimum variance) & $0.172$ & $0.138$ & $0.198$\\
$\Neff$ of $q_{\mathrm{aff}}$                          & $33.8$  & $25.5$  & $52.7$\\
$\Neff$ (minimum variance on $\Del$)                  & $28.9$  & $18.8$  & $49.6$\\
support size of the minimum-variance portfolio        & $52/255$& $40$    & $80$\\
condition number of $\Sigh$                           & $224$   & $115$   & $651$\\
$\lambda_{\min}(\Sigh)$                               & $1.43\times10^{-4}$ & $7.8\times10^{-5}$ & $2.6\times10^{-4}$\\
$\kappa_\Delta$                                       & $1.43\times10^{-4}$ & $7.8\times10^{-5}$ & $2.6\times10^{-4}$\\
$\kappa_\Delta/\lambda_{\min}(\Sigh)$                  & $1.00$  & $1.00$  & $1.00$\\
\hline
\end{tabular}
\end{table}

Two facts in Table~\ref{tab:ashare} deserve emphasis. First, the benchmark level
of breadth is low: the long-only minimum-variance portfolio holds only about
fifty-two names out of $255$ in a typical window and has an effective number of
about twenty-nine, and even the affine solution $q_{\mathrm{aff}}$, which is
free to short, reaches only $33.8$. Either way $\Neff$ is an order of magnitude
below $N$. Risk heterogeneity is therefore economically large, and a
breadth discussion is not an academic curiosity. Second, restricting the
curvature constant to the feasible direction set changes nothing: the ratio
$\kappa_\Delta/\lambda_{\min}(\Sigh)$ equals one in all $140$ windows. We
verified this on real data before using it in a theorem, and report it as a
negative result: tightening a curvature bound by restricting to the tangent
space of the simplex is not a source of improvement, and any paper claiming such
an improvement should exhibit the ratio.

\section{The radius is a breadth parameter}\label{sec:breadth}

\begin{proposition}[Endogenous breadth cap]\label{prop:cap}
Let $C$ be nonempty, $g\ge0$, $\tau\ge0$, and let $\phid^\star$ be a global
maximizer of $U(\phid)-\tau g(\phid)$ over $C$, which we assume exists. Put
$h^\star=g(\phid^\star)$. Then $\phid^\star$ is \emph{an} optimal solution of
$\max\{U(\phid):\phid\in C,\ g(\phid)\le h^\star\}$.
\end{proposition}

\begin{proof}
For any $\phid\in C$ with $g(\phid)\le h^\star$, optimality of $\phid^\star$ gives
$U(\phid)-\tau g(\phid)\le U(\phid^\star)-\tau h^\star$, hence
$U(\phid)\le U(\phid^\star)+\tau(g(\phid)-h^\star)\le U(\phid^\star)$.
\end{proof}

\begin{remark}
The statement is a supporting-hyperplane property, not an equivalence of
optimization problems: the constrained problem may have additional optimizers
which are not optimizers of the penalized problem. What it does say is that the
penalized solution can always be described without any reference to
distributional uncertainty.
\end{remark}

On the long-only simplex with $p=2$ we have $\norm{\phid}_2^2=\HHI(\phid)$, so
the penalized problem reads
\begin{equation}\label{eq:hhi}
  \max_{\phid\in\Del}\ U(\phid)-\tau\sqrt{\HHI(\phid)},
\end{equation}
and the marginal rate of substitution between utility and effective breadth is
$\tau=\sqrt{\delta}$: the price applies to $\sqrt{\HHI}$, not to $\HHI$. We
therefore refer to $\tau$ as the price of breadth. With $p=1$ the penalty is
\emph{identically constant} on $\Del$ because $\norm{\phid}_1=\mathbf{1}'\phid=1$;
$(P)$ then coincides with ordinary mean--variance, while the constraint of $(K)$
reduces to the scalar condition $\sqrt{\delta}\le\max_i\muh_i-r_0$, which for
$r_0=\mathbf{1}'\muh/N$ becomes $a:=\sqrt{\delta}/s(\muh)\le1$, where
$s(\muh)=\max_i\muh_i-\mathbf{1}'\muh/N>0$; the ratio is defined only when
$s(\muh)>0$, and if $s(\muh)=0$ the condition is $\sqrt{\delta}\le0$. The two values of $p$ are therefore not
interchangeable in the long-only problem: one leaves a constant, the other a
breadth price.

\begin{figure}[t]
\centering
\includegraphics[width=0.55\textwidth]{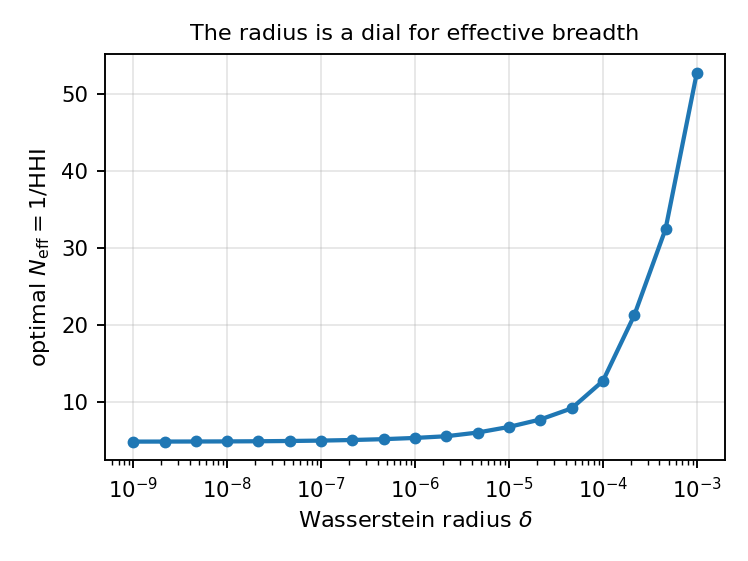}
\caption{The radius is a dial for breadth, as an empirical description rather than as a
statement of a one-to-one mapping. Optimal effective number of holdings
$\Neff=1/\HHI$ of the penalized robust portfolio as a function of the radius
$\delta$, on simulated data with $N=255$, $T=60$, $\lambda=10$. The map is
monotone by Proposition~\ref{prop:monotone} and, within the empirically relevant
range, moves $\Neff$ from six to a hundred and forty.}\label{fig:breadth}
\end{figure}

\begin{figure}[t]
\centering
\includegraphics[width=0.55\textwidth]{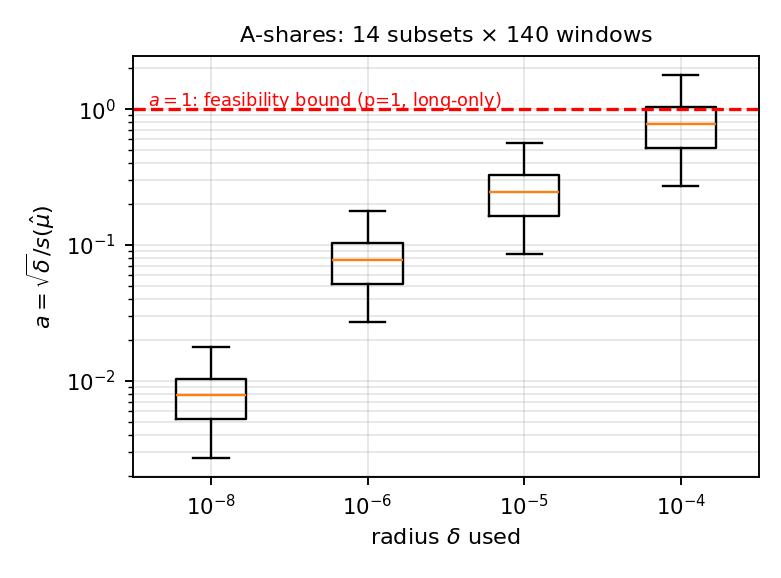}
\caption{Calibration of the radius on fourteen A-share panels and $140$ windows.
Boxes show the distribution of the dimensionless ratio
$a=\sqrt{\delta}/s(\muh)$ for four values of $\delta$; the dashed line is
$a=1$, which is the feasibility boundary of the return-floored problem with
$p=1$ and $r_0=\mathbf{1}'\muh/N$.}\label{fig:acalib}
\end{figure}

Finally, the complexity path is monotone.

\begin{proposition}[Monotone complexity path]\label{prop:monotone}
Let $g\ge0$, $\tau_2>\tau_1\ge0$, and let $\phid_1,\phid_2$ be any global
maximizers of $U-\tau_1g$ and $U-\tau_2g$ over the same $C$. Then
$g(\phid_2)\le g(\phid_1)$.
\end{proposition}

\begin{proof}
Add the two optimality inequalities
$U(\phid_1)-\tau_1g(\phid_1)\ge U(\phid_2)-\tau_1g(\phid_2)$ and
$U(\phid_2)-\tau_2g(\phid_2)\ge U(\phid_1)-\tau_2g(\phid_1)$.
\end{proof}

The conclusion is only monotonicity. It does not give uniqueness, continuity, or
a one-to-one map from $\tau$ to breadth, and in particular it does not identify
$\tau$ from observed weights; flat regions of $U$ or locally constant penalties
break strict monotonicity. Since $\norm{\phid}_2$ attains its minimum
$N^{-1/2}$ uniquely at $\phid=\mathbf{1}/N$, the limiting behaviour of $(P)$ as
$\tau\to\infty$ is the equally weighted portfolio \emph{when $C$ is the full
simplex}; for a smaller $C$ the limit is the minimum-norm point of $C$, and we do
not claim otherwise.

\section{Identification}\label{sec:identification}

\subsection{Observational equivalence}

\begin{proposition}[Observational equivalence]\label{prop:obs}
Let $X$ denote training data, $Z$ test data and $Y$ test-period returns. Two
methods generate weight processes $W^{A}_t=G_A(X,Z_{1:t-1})$ and
$W^{B}_t=G_B(X,Z_{1:t-1})$. If $W^{A}_t=W^{B}_t$ almost surely for every $t$, and
the two methods use the same trading rule, rebalancing dates, transaction costs,
prices and return data, then any statistic $S=H(X,Z,Y,W)$ constructed from these
observables satisfies $S^{A}=S^{B}$ almost surely.
\end{proposition}

\begin{proof}
The weight paths coincide, so every trade and every realized return coincides
pathwise; all statistics built from them are therefore equal.
\end{proof}

What the lemma does \emph{not} say is equally important: it does not imply that
the internal optimization problems agree, nor that predictions coincide out of
sample, nor that robustness is theoretically vacuous. It says that, given the
reported observables, a study cannot attribute an improvement to robustness
rather than to an equivalent shrinkage or constraint.

\subsection{The breadth-matched benchmark}

\begin{theorem}[Endogenous breadth cap]\label{thm:identity}
Let $C$ be a nonempty closed convex subset of the simplex $\Del$, let
$\lambda>0$, let $\Sigh$ be positive semidefinite and let $\tau>0$. If
$\phid^\star_\delta$ solves \eqref{eq:P} with $p=2$, and
$h^\star=\norm{\phid^\star_\delta}_2$, then $\phid^\star_\delta$ is the
\emph{unique} solution of
\begin{equation}\label{eq:ball}
  \max\bigl\{\, U(\phid)\;:\;\phid\in C,\ \norm{\phid}_2\le h^\star \,\bigr\}.
\end{equation}
In particular $U(\phid^\star_\delta)=\max\{\cdots\}$ and the breadth constraint in
\eqref{eq:ball} is active at the optimum.
\end{theorem}

\begin{proof}
Let $\phid$ be feasible for \eqref{eq:ball}. Optimality of $\phid^\star_\delta$ in
\eqref{eq:P} gives
$U(\phid)-\tau\norm{\phid}_2\le U(\phid^\star_\delta)-\tau h^\star$, hence
$U(\phid)\le U(\phid^\star_\delta)+\tau(\norm{\phid}_2-h^\star)\le U(\phid^\star_\delta)$,
so $\phid^\star_\delta$ is optimal and the optimal values agree. Now let $\phid$
be any optimal solution of \eqref{eq:ball}. Then $U(\phid)=U(\phid^\star_\delta)$,
and substituting into the same inequality gives
$\tau\norm{\phid}_2\ge \tau h^\star$, hence $\norm{\phid}_2=h^\star$ because
$\phid$ is feasible. Suppose $\phid\neq\phid^\star_\delta$ and put
$\bar\phid=(\phid+\phid^\star_\delta)/2$, which lies in $C$ by convexity. Two
distinct points of the simplex are not positive scalar multiples of one another,
so strict convexity of the Euclidean norm gives
$\norm{\bar\phid}_2<(\norm{\phid}_2+\norm{\phid^\star_\delta}_2)/2=h^\star$.
Concavity of $U$ gives $U(\bar\phid)\ge (U(\phid)+U(\phid^\star_\delta))/2=U(\phid^\star_\delta)$,
and therefore
\[
  U(\bar\phid)-\tau\norm{\bar\phid}_2 \;>\; U(\phid^\star_\delta)-\tau h^\star,
\]
contradicting the optimality of $\phid^\star_\delta$ in \eqref{eq:P}. Hence
$\phid=\phid^\star_\delta$.
\end{proof}

\begin{remark}[What the theorem does and does not need]
Existence follows from $C$ compact, and $\Sigh\succ0$ is \emph{not} required: the
norm term supplies the strict concavity that the argument uses, through the
observation that distinct simplex points cannot lie on a common ray. Position
caps and active nonnegativity constraints are absorbed into the normal cone of
$C$ and do not affect the argument, which is global and uses no differentiability.
The conclusion fails at $\tau=0$, where \eqref{eq:P} is ordinary mean--variance
selection and multiple optima can occur.
\end{remark}

\begin{corollary}[Multiplier correspondence]\label{cor:multiplier}
Let $h^\star=\norm{\phid^\star_\delta}_2$. If \eqref{eq:ball} is written with the
squared constraint $\norm{\phid}_2^2\le (h^\star)^2$, its radial multiplier is
$\nu=\tau/(2h^\star)$; if it is written with $\norm{\phid}_2\le h^\star$, its
multiplier is $\tau$. The correspondence holds with active nonnegativity
constraints, active caps and the simplex equality, since the normal-cone terms are
common to both stationarity systems and only the radial term differs. No
constraint qualification on $C$ is required for the radial value itself: the
first-order condition of \eqref{eq:P} already supplies a normal vector of $C$ at
$\phid^\star$, and the statement is that the corresponding stationarity condition
holds with $\nu=\tau/(2h^\star)$. Individual multipliers for the nonnegativity and
cap constraints may require a qualification such as LICQ, which the comparison
above does not.
\end{corollary}

The corollary is what makes the identity operational: the breadth-matched
portfolio is obtained from a single mean--variance problem in which the
covariance penalty is $\phid'(\lambda\Sigh+\nu I)\phid$ with
$\nu=\tau/(2h^\star)$, so no search over the constraint is needed. This is a
ridge penalty on the weights, and it coincides with raising the risk-aversion
parameter only when $\Sigh$ is a multiple of the identity; we therefore do not
describe it as a change in risk aversion. Section~\ref{sec:evidence} uses the
corollary in this form, and reports the residual
$\norm{\phid^{\mathrm{cap}}(h^\star)-\phid^\star_\delta}_2$ as a check on the
solvers.

\begin{proposition}[Scope of the equivalence]\label{prop:scope}
The equivalence of Theorem~\ref{thm:identity} is a property of the penalised
scalarisation \eqref{eq:P}. Nothing in that argument transfers to the
return-floored problem \eqref{eq:K} or to the exact robust problem \eqref{eq:D},
and the following two arguments show that the corresponding statements are false
or unavailable, in that order.
\end{proposition}

\begin{proof}[The statement is false for \eqref{eq:K}]
Take $N=2$, $C=\Del$, $\lambda=1$,
\[
  \mu=(0,2)',\qquad r_0=\mathbf{1}'\mu/N=1,\qquad \tau=\tfrac12,\qquad
  \Sigma=\mathrm{diag}(1,3).
\]
Write $\phid(x)=(1-x,x)$. Then
$U(\phid(x))=2x-\bigl((1-x)^2+3x^2\bigr)=-1+4x-4x^2$, which is uniquely maximised
at $x=\tfrac12$. The feasibility condition of \eqref{eq:K} is
\[
  2x-\tfrac12\sqrt{(1-x)^2+x^2}\ \ge\ 1 .
\]
At $x=\tfrac12$ the left-hand side equals $1-\tfrac{1}{2\sqrt2}<1$, and at $x=1$
it equals $\tfrac32$, so the feasible set is $[x_0,1]$ for some $x_0>\tfrac12$.
Since $U$ is strictly decreasing on $(\tfrac12,1]$, the unique optimizer of
\eqref{eq:K} is $\phid(x_0)$. But $\norm{\phid(1/2)}_2<\norm{\phid(x_0)}_2$ while
$U(\phid(1/2))>U(\phid(x_0))$, so $\phid(x_0)$ does not solve
$\max\{U(\phid):\phid\in\Del,\ \norm{\phid}_2\le\norm{\phid(x_0)}_2\}$. The radius
in \eqref{eq:K} changes feasibility through the return term and through the norm
term at the same time, and the resulting feasible set is not a Euclidean ball.
\end{proof}

\begin{proof}[The argument is unavailable for \eqref{eq:D}]
Proposition~\ref{prop:cap} is a statement about objectives of the form
$U(\phid)-\tau g(\phid)$. The objective of \eqref{eq:D} is a square of a sum of a
standard deviation and a multiple of the norm, so the supporting-hyperplane
argument cannot be invoked, and the solution of \eqref{eq:D} is not, in general,
characterised by a single radial first-order term. We therefore claim only that
the equivalence is not implied; we do not claim that no equivalence of any other
form could exist. The feasible set of \eqref{eq:D} does shrink monotonically in
$\delta$ under its own definition, but monotonicity of a feasible set is not a
breadth-cap representation.
\end{proof}

We therefore state the scope as a non-implication rather than as an impossibility:
the endogenous-breadth-cap result holds for the penalised scalarisation, and it
does not extend to the other two. Section~\ref{sec:evidence} reports the three
portfolios side by side, which is the empirical counterpart of this restriction.

\subsection{The three scalarizations are not interchangeable in practice}

Proposition~\ref{prop:scope} is not only a logical point. We solved all three
problems on the same windows, and they select very different books.
Table~\ref{tab:threeforms} reports the comparison. The penalised problem holds a
book with median Euclidean norm 0.38, about seven effective holdings; the exact
robust problem holds one with norm 0.18, about thirty-one holdings; and the
return-floored problem is driven to a corner, with norm 0.88 and about one
effective holding. Out of sample the ordering against the equally weighted
portfolio follows the concentration: $+57.5$ basis points per day for the
return-floored form, $+45.7$ for the penalised form and $+10.8$ for the exact
robust form at $\delta=10^{-6}$. An analysis that treats the three as
interchangeable would therefore conflate effects of very different sign and
magnitude.

\begin{table}[t]
\centering
\caption{The three scalarisations on the same sixty-day windows. Breadth is
summarised by the median Euclidean norm of the portfolio and by the implied
effective number of holdings; slack is the value of the left-hand side minus the
right-hand side of the corresponding return constraint, evaluated at the solution;
the last two columns give the Euclidean distance to the penalised portfolio and
the out-of-sample gain against the equally weighted portfolio at $\delta=10^{-6}$,
in units of $10^{-4}$ per day. Six panels, sixty-day estimation window, ten-day
holding period.}\label{tab:threeforms}
\begin{tabular}{lcccc}
\hline
 & median $\norm{\phid}_2$ & implied $\Neff$ & slack & gain vs EW\\
\hline
penalised $(P)$ & $0.375$ & $7.1$ & -- & $+45.7$\\
exact $(D)$ & $0.179$ & $31.2$ & $1.7\times10^{-3}$ & $+10.8$\\
return-floored $(K)$ & $0.882$ & $1.3$ & $1.2\times10^{-2}$ & $+57.5$\\
\hline
\multicolumn{5}{l}{$\norm{\phid^{(P)}-\phid^{(D)}}=0.34$;
$\norm{\phid^{(P)}-\phid^{(K)}}=0.53$; robust versus breadth-matched residual
$1.3\times10^{-5}$ (median)}\\
\hline
\end{tabular}
\end{table}

The last line of the table records the accuracy of the identity: solving the
breadth-matched problem at the penalised portfolio's own norm reproduces it to a
median Euclidean distance of $1.3\times10^{-5}$, and the separately implemented
norm-cap solver hits its target norm to $8\times10^{-9}$. The identity of
Theorem~\ref{thm:identity} is therefore satisfied by our solvers at the precision
we report, which is the sense in which we treat it as a check rather than as
evidence.

\subsection{The weights are not identified}

Theorem~\ref{thm:identity} pins down values, not weights. If the constrained
maximizer fails to be unique, or if the two solutions are computed to finite
precision, the portfolios may differ, and the difference is confined to the
eigen-directions of $\Sigh$ on which the objective is nearly flat. The curvature
of $\Sigh$ therefore plays a second role in this paper, distinct from any
breadth bound: it governs whether the effect of the radius on the weights is
identifiable at all. In the numerical experiments reported below, the residual
weight difference ranges from $10^{-8}$ to $5\times10^{-3}$ while the objective
difference stays at the level of solver tolerance; a statistic built on realized
returns inherits the weight difference, not the objective difference.

This has a direct consequence for inference. When two portfolios almost coincide,
the difference of their return series has almost no variance, so any small
systematic solver bias produces a large $t$-statistic. Significance alone is
therefore an unreliable criterion for the comparison against the breadth-matched
benchmark; the appropriate criterion is economic magnitude alongside the
numerically attainable resolution, which we report explicitly in
Section~\ref{sec:evidence}.

\subsection{What an empirical study must report}

The audit of Table~\ref{tab:audit} is the minimal version of the exercise and
takes one extra column. Collecting the above, a study that wishes to attribute an
improvement to robustness should report, at a minimum: (i) the equally weighted portfolio;
(ii) ordinary mean--variance portfolios under at least two covariance estimators;
(iii) the breadth-matched benchmark of Theorem~\ref{thm:identity}, matched on
$\Neff$ rather than on the radius; (iv) a direct norm-constrained portfolio, that
is \eqref{eq:P} with $h$ on a grid rather than $\delta$; (v) the return-floored
problem \eqref{eq:K}; (vi) shrinkage estimators, both in the mean and towards the
equally weighted portfolio; (vii) turnover, cost and exposure on a common basis;
(viii) a pre-specified rule for $\delta$; (ix) $\lambda_{\min}(\Sigh)$ and
$\norm{q}_2$, because Theorem~\ref{thm:impossibility} and
Section~\ref{sec:identification} show that breadth and identifiability are
governed by them; and (x) pointwise similarity of the weight maps, not only
terminal performance.

\section{Evidence}\label{sec:evidence}

\subsection{Data, design and structural quantities}

Data are fourteen A-share panels of $N=255$ daily returns, each covering 648
trading days between March 2018 and January 2024, taken from the public
Stockformer dataset. Portfolios are estimated on a 60-day rolling window and held
for five days, so the out-of-sample return blocks are disjoint within a panel.
The covariance matrix is Ledoit--Wolf \citep{ledoit2004} unless stated otherwise,
expected returns are sample means, and $\lambda=10$ unless stated otherwise.

The panels themselves, with a manifest of per-file SHA-256 checksums, and the code
that produces every table and figure of this section are openly available at
\url{https://huggingface.co/datasets/Changahou/GU2026_1} and
\url{https://github.com/ChenJiahaoJNU/GU2026-11}; both are cited again in the
availability statement at the end of the paper. What is redistributed is the
daily return matrix alone: no prices, identifiers, volumes or factor values.

Four benchmarks are compared with the robust portfolio
$\phid^{\mathrm{DRO}}_\delta$: the breadth-matched portfolio of
Theorem~\ref{thm:identity} with $\nu=\sqrt{\delta}/(2h^\star)$; a return-matched
portfolio solving \eqref{eq:K} with $r_0=\mathbf{1}'\muh/N$; ordinary
mean--variance with the same $\lambda$; and the equally weighted portfolio. We
refer to the resulting out-of-sample average daily return differences as
$\mathrm{Rent}(\cdot)$.

Two features of the inference should be stated at the outset. First, the
comparisons are numerous: each panel contributes five tracking rules at two radii,
so ten contrasts against the robust portfolio. We therefore adjust by a
Romano--Wolf step-down procedure \citep{romano2005} applied within each panel, with
the null distribution of the maximum studentised statistic obtained from a
stationary block bootstrap, and we report adjusted rather than raw p-values. The
residual of every tracking rule survives the correction: the median adjusted
p-value is between $0.009$ and $0.025$ across the ten contrasts, and between five
and six of the six panels remain significant at the five per cent level for each
contrast. The conclusion is stable in the block length, at twenty and at forty
days. A caveat belongs with these numbers: in several cells the bootstrap-adjusted
p-value is smaller than the analytic one, which means the two variance estimators
disagree, and we therefore read the multiplicity conclusion as a count of surviving
panels rather than as a precise p-value. Second, the panels overlap in time.

Inference uses Newey--West heteroskedasticity and autocorrelation consistent
standard errors \citep{newey1987} with lag $\max(4\,\mathrm{HOLD},10)$ and a
stationary bootstrap \citep{politis1994} with $B=2000$ resamples and expected
block length $20$, applied to the continuous daily series rather than to the
fourteen overlapping panels. Because the fourteen panels are rolling slices of
the same market, we do not pool them for significance testing; we report
consistency counts, and we additionally report two \emph{disjoint} panels
(2018-03 to 2020-10 and 2021-06 to 2024-01) as separate samples.

Table~\ref{tab:ashare} already summarizes the structural quantities. Two further
diagnostics are informative for interpreting the bound in Appendix~\ref{app:bound}.
The mean curvature along feasible directions,
$\mathrm{tr}(\Sigma_{1^\perp})/(N-1)$, exceeds $\lambda_{\min}(\Sigh)$ by a
median factor of $4.3$, and the Rayleigh quotient in the direction of the
estimate error, $\phid^{EW}-\phid^{\star}$, exceeds it by a median factor of
$15.5$. The looseness of any bound built on $\lambda_{\min}$ is therefore
attributable to the difference between the worst direction and the realized
direction, not to the feasible set, since
$\kappa_\Delta/\lambda_{\min}=1$.

\subsection{The rent under four benchmarks}

\begin{table}[t]
\centering
\caption{Rent of distributional robustness relative to four benchmarks, in units
of $10^{-4}$ average daily return (positive means the robust portfolio is
better), fourteen A-share panels. ``sig'' counts panels in which the
Newey--West $t$-statistic exceeds two in absolute value and the stationary
bootstrap interval excludes zero. The last column reports the median absolute
in-sample difference in the mean--variance objective between the robust
portfolio and the breadth-matched benchmark.}\label{tab:rent}
\begin{tabular}{lcccccccc}
\hline
 & \multicolumn{2}{c}{breadth-matched} & \multicolumn{2}{c}{return-matched}
 & \multicolumn{2}{c}{ordinary MV} & \multicolumn{2}{c}{equally weighted}\\
\cline{2-3}\cline{4-5}\cline{6-7}\cline{8-9}
$\delta$ & rent & sig & rent & sig & rent & sig & rent & sig\\
\hline
$10^{-6}$ & $+0.01$ & $0/14$ & $-2.05$ & $0/14$ & $-0.60$ & $1/14$ & $+29.82$ & $7/14$\\
$10^{-5}$ & $+0.00$ & $0/14$ & $-2.62$ & $1/14$ & $-1.15$ & $1/14$ & $+29.25$ & $7/14$\\
$10^{-4}$ & $+0.00$ & $0/14$ & $-4.75$ & $1/14$ & $-5.27$ & $4/14$ & $+24.63$ & $7/14$\\
\hline
\multicolumn{9}{l}{median $|U(\phid^{B})-U(\phid^{\star}_\delta)|$:
$1.8\times10^{-8}$, $1.2\times10^{-9}$, $8.6\times10^{-15}$ for the three radii}\\
\hline
\end{tabular}
\end{table}

Table~\ref{tab:rent} reports the four ex post comparisons over all fourteen
panels. The breadth-matched column is the accuracy check discussed above: it is
zero by Theorem~\ref{thm:identity}, and the in-sample objective difference falls to
$10^{-15}$ as the radius grows, which is what we want from the solvers rather than
from the data. The informative columns are the other three. Against the
return-matched benchmark the rent is negative; against ordinary mean--variance it
is negative and significant in four of fourteen panels at $\delta=10^{-4}$; against
the equally weighted portfolio it is large and positive, and significant in seven
of fourteen panels.

The reading is that the sign and the significance of the reported effect are set by
the benchmark, so that a study reporting only one of these comparisons is reporting
a number whose interpretation depends on a choice it has not disclosed. The large
positive figure against the equally weighted portfolio measures the value of risk
optimization and breadth; the negative figures measure the return cost of the
penalty. Neither is a measurement of robustness, and the ex ante test of
Table~\ref{tab:audit} is where the question of robustness is actually answered.

\subsection{Robustness of the finding}

Two sample designs are used in this section and they should not be confused. The
four-benchmark comparison of Table~\ref{tab:rent} and the multi-setting check that
follows use all fourteen panels with a five-day holding period. The ex ante audit
of Table~\ref{tab:audit} and the three-form comparison of
Table~\ref{tab:threeforms} use the first six panels with a ten-day holding period,
because those exercises require several constrained solves per window. Levels are
therefore not comparable across the two designs; the comparisons within each are.

The finding was checked across eight settings: $\lambda\in\{30,100\}$;
$\Sigh$ from OAS and from the (rank-deficient) sample covariance; rebalancing
horizons of one and twenty days; and position caps of $0.10$ and $0.05$. In every
setting the breadth-matched rent has median absolute value at most
$0.01\times10^{-4}$ per day and is at most $0.1\%$ of the corresponding
return-matched or equally weighted effect. Three observations are worth recording.
First, using the rank-deficient sample covariance leaves the conclusion unchanged,
which is consistent with Theorem~\ref{thm:identity} depending only on the pair
$(U,C)$: non-uniqueness of the optimizer increases the weight difference without
affecting the value identity. Second, tightening the position cap compresses all
effects: the equally weighted rent falls from about $+29\times10^{-4}$ per day
without caps to $+8.7$ with a $0.10$ cap and $+5.7$ with a $0.05$ cap. Third, at
$\lambda=30$ the breadth-matched rent shows twelve significant panels despite a
median magnitude of $-0.009\times10^{-4}$ per day; in the same batch the
numerical objective difference is largest ($5.9\times10^{-8}$ against
$10^{-14}$ in the baseline), which identifies the significance as solver
artefact through the near-zero-variance mechanism of
Section~\ref{sec:identification}. Transaction costs and two disjoint subperiods
are examined next.

Because the fourteen panels overlap in time, we repeat the exercise on two
\emph{disjoint} continuous panels and add turnover costs of thirty basis points,
charged once per rebalance on the $\ell_1$ turnover and amortized over the
five-day holding period. Table~\ref{tab:periods} reports the result.

\begin{table}[t]
\centering
\caption{Two disjoint subperiods, $\lambda=10$, in units of $10^{-4}$ average
daily return (Newey--West $t$-statistics in parentheses). The breadth-matched
rent stays at zero in both periods, whereas the rents against the
return-matched benchmark and against the equally weighted portfolio change sign
between periods. The last column adds costs of thirty basis points.}
\label{tab:periods}
\begin{tabular}{lcccc}
\hline
 & \multicolumn{3}{c}{panel A: 2018-03 to 2020-10} & panel B: 2021-06 to 2024-01\\
\cline{2-4}
$\delta$ & breadth-matched & return-matched & equally weighted &
breadth-matched, net of costs\\
\hline
$10^{-6}$ & $-0.013$ ($-0.79$) & $-21.08$ ($-1.73$) & $+50.33$ ($+2.60$) & $-0.010$ ($-0.64$)\\
$10^{-5}$ & $-0.017$ ($-1.33$) & $-24.54$ ($-1.88$) & $+46.86$ ($+2.61$) & $-0.016$ ($-1.25$)\\
$10^{-4}$ & $-0.001$ ($-1.59$) & $-36.05$ ($-2.16$) & $+35.36$ ($+2.57$) & $+0.002$ ($+1.19$)\\
$10^{-3}$ & $-0.000$ ($-0.18$) & $-59.25$ ($-2.52$) & $+12.16$ ($+2.20$) & $+0.000$ ($+1.85$)\\
\hline
 & \multicolumn{3}{c}{panel B: 2021-06 to 2024-01} & panel B, net of costs\\
\cline{2-4}
$10^{-6}$ & $+0.010$ ($+1.62$) & $+0.69$ ($+0.05$) & $-4.61$ ($-0.67$) & $+0.011$ ($+1.69$)\\
$10^{-5}$ & $+0.003$ ($+1.38$) & $+0.57$ ($+0.04$) & $-4.73$ ($-0.74$) & $+0.004$ ($+1.81$)\\
$10^{-4}$ & $+0.000$ ($+0.94$) & $+1.27$ ($+0.08$) & $-4.03$ ($-0.78$) & $+0.001$ ($+1.90$)\\
$10^{-3}$ & $-0.000$ ($-0.01$) & $+1.80$ ($+0.10$) & $-3.50$ ($-1.24$) & $+0.000$ ($+1.30$)\\
\hline
\multicolumn{5}{l}{median $|U(\phid^{B})-U(\phid^{\star}_\delta)|$: $9.6\times10^{-9}$ (A), $5.6\times10^{-9}$ (B)}\\
\hline
\end{tabular}
\end{table}

Two conclusions follow. First, the breadth-matched rent is again at the level of
solver precision in both subperiods and is not significant at $\lambda=10$ for
any radius, while the numerical objective gap is $10^{-8}$ or smaller; turnover
costs of thirty basis points do not change this, as expected since the two
portfolios have almost identical turnover. Second, and more importantly for the
interpretation of the literature, the rents against the two naive benchmarks
\emph{change sign} between subperiods: against the equally weighted portfolio the
robust portfolio gains $+50\times10^{-4}$ per day in panel A and loses
$-4.6\times10^{-4}$ in panel B. A comparison that is reported without a
breadth-matched benchmark is therefore not merely uninformative about robustness;
it is not even a stable statement about which strategy is better.

\subsection{Known-truth simulation}

To verify that the mechanism is not an artefact of the empirical design, we
simulate from a known factor model, $r_t=\mu+Bf_t+\varepsilon_t$ with five
factors and a sparse alpha of magnitude $c\,\sigma_i/\sqrt{T}$ on twenty percent
of assets, and we evaluate the mean--variance regret
$U(\phid^\star)-U(\widehat\phid)$ against the true optimum. With $N=255$,
$T=60$, $\lambda=10$ and $300$ replications, the ordering of median regret as a
percentage of the attainable utility is: breadth-matched minimum variance and
exactly solved robust portfolios $83\%$; equally weighted $91\%$; robust plus
equally weighted $197\%$; mean shrinkage $270\%$; penalized robust with a wide
radius $296\%$; penalized robust with a validation-selected radius $471\%$;
sample covariance mean--variance $548\%$. The increment of the robustness term
over the breadth-matched benchmark is $-1.9\times10^{-6}$ per day, against
$-6.1\times10^{-4}$ for the breadth tilt and $+1.3\times10^{-3}$ for the
mean--variance penalty relative to the equally weighted portfolio: a factor of
three hundred. With heavy-tailed innovations drawn from a $t_3$ distribution
(rescaled to unit variance) the ordering is unchanged in all sixteen
configurations of the design grid: the breadth-matched minimum-variance
benchmark and the exactly solved robust portfolio are better than the equally
weighted portfolio in sixteen of sixteen configurations, attaining on average
$83\%$--$86\%$ of the attainable utility at the headline setting, while sample
mean--variance attains a regret of $167\%$--$1574\%$ depending on $N$, $T$ and
$\lambda$. The penalty paid by the robustness term relative to the
return-matched benchmark remains at the level of $10^{-6}$ per day in every
configuration, three orders of magnitude below the effects that separate the
competing strategies. Imposing a risk--return trade-off through
$\mu\propto\sigma$ also leaves the ordering intact while lowering the level of
regret. Note that the synthetic experiment compares against the return-matched
benchmark, not the breadth-matched one; the breadth-matched identity is verified
directly on the A-share panels in Table~\ref{tab:rent}, where the objective gap is
at most $10^{-8}$ and typically $10^{-13}$.

\section{Conclusions}\label{sec:discussion}

\subsection*{Findings}

The Wasserstein radius in the long-only mean--variance problem is a breadth
parameter. Because the robustification term depends on the portfolio only through
its norm, the penalised robust solution is an ordinary mean--variance solution
subject to an endogenous cap on breadth, and on the simplex the multiplier of that
cap prices the effective number of holdings. Theorem~\ref{thm:identity} states
this as an equality of optimisers, with uniqueness and without a positive
definiteness assumption, and Corollary~\ref{cor:multiplier} converts it into a
single multiplier, so the cap can be imposed in one solve rather than searched for.

Two consequences follow. First, no breadth guarantee follows from the spectral
properties of the covariance matrix: with a strictly positive smallest eigenvalue
and zero cross-sectional alpha, the long-only optimum can approach a single asset,
so the breadth of the minimum-variance benchmark must be an explicit input to any
such guarantee. Second, for the penalised scalarisation the radius carries no
in-sample information beyond the choice of breadth. That equivalence is confined
to the penalised scalarisation: the return-floored and the exact robust problems
are different problems, and on our data they select books with about seven,
thirty-one and one effective holdings respectively.

Because the ex post comparison is exact by construction, the informative empirical
question is what an ex ante cap can recover. Table~\ref{tab:audit} answers it: a
cap that tracks the past level of the robust portfolio's realised breadth recovers
about five sixths of its gain, and the residual of seven to eleven basis points
per day is stable across five tracking rules and survives a Romano--Wolf
correction in five or six of six panels, whereas a cap fixed at an absolute target
recovers only a quarter to two fifths. The robust portfolio is therefore best
described as a breadth rule, whose realised level responds to the current window,
rather than as a robustness premium or as a single fixed breadth.

\subsection*{Implications}

For practice, the implication is that calibrating the radius carefully while
leaving breadth to the calibration is tuning the smaller of two parameters. The
quantity worth choosing, and worth reporting, is the effective number of holdings
or the price the manager is willing to pay for it, and the choice can be made
directly through a norm constraint without invoking an ambiguity set. For the
literature, the implication is about attribution rather than about method. A
statement that a robust portfolio outperforms a sample-based one is consistent
with the radius having no effect at all, because the two portfolios differ in
breadth. The reporting standard of Section~\ref{sec:identification} is designed to
make the difference visible, and it is cheap to satisfy: the breadth-matched
benchmark and the implied effective number of holdings are two additional columns
in a table that already exists. For theory, the implication is that results which
guarantee the value of a robust decision are not, by themselves, results about
the composition of the portfolio. Closing that gap requires either a curvature
condition on the feasible directions, as in Appendix~\ref{app:bound}, or a
different formulation in which the norm is not a breadth measure, such as the
multi-asset setting with short sales.

\subsection*{Limitations and outlook}

The impossibility theorem uses a diagonal covariance witness and establishes that
no uniform guarantee exists; it does not show that breadth guarantees are
unattainable under economically plausible structures, and characterizing the
structures under which they are attainable is open. The identification theorem is
static: with transaction costs or multiple periods the feasible set changes and
the argument must be re-derived, although the supporting-hyperplane property that
it rests on holds for any convex feasible set. The empirical work covers a single
market and a single asset universe of 255 names, and because the fourteen panels
overlap in time, the inference rests on continuous series within panels and on two
disjoint subperiods rather than on cross-panel independence. Finally, we cannot
offer a distribution-free breadth bound, and Appendix~\ref{app:bound} explains
why: with the sample covariance the relevant curvature is zero and any bound of
that form is vacuous, so a guarantee is available only in combination with a
covariance regularizer whose spectral floor can be quantified. That combination,
rather than the radius itself, is where the remaining theoretical work lies.

\appendix

\section{A conditional breadth bound}\label{app:bound}

For completeness we record what can be said with additional structure, and we
begin with the point that makes the statement delicate: the covariance matrix in
the curvature constant must be the one used to define the objective whose gap is
measured. Two statements are available, and they are not the same statement.

\emph{Population statement.} Let
$U_\Sigma(\phid)=\phid'\mu-\lambda\phid'\Sigma\phid$ with $\lambda>0$ and
$\Sigma\succ0$, and let $\phid^\star$ maximise $U_\Sigma$ over a convex set $C$.
If $\widehat\phid\in C$ satisfies
$U_\Sigma(\phid^\star)-U_\Sigma(\widehat\phid)\le\epsilon$, then the
first-order condition at $\phid^\star$ and strong concavity with modulus
$2\lambda\lambda_{\min}(\Sigma)$ give
\begin{equation}\label{eq:growth}
  U_\Sigma(\phid^\star)-U_\Sigma(\widehat\phid)\ \ge\
  \lambda\,\lambda_{\min}(\Sigma)\,\norm{\widehat\phid-\phid^\star}_2^2,
\end{equation}
for every convex $C$, including sets with active constraints and corners; the
inequality is the statement we use and boundary optima need no separate
treatment. Consequently
\begin{equation}\label{eq:nbound}
  \Neff(\widehat\phid)\ \ge\ \Bigl(\norm{\phid^\star}_2+
  \sqrt{\epsilon/(\lambda\lambda_{\min}(\Sigma))}\Bigr)^{-2}.
\end{equation}
Here $\lambda_{\min}(\Sigma)$ is generally unknown, and it cannot be replaced by
an estimator without a statistical argument.

\emph{Estimated statement.} If instead $\epsilon$ denotes the gap in the sample
objective $\widehat U(\phid)=\phid'\muh-\lambda\phid'\Sigh\phid$, the same
argument gives \eqref{eq:nbound} with $\Sigh$ and $\phid^\star$ replaced by the
sample optimizer. This is the version for which numbers can be reported. With
$N=255$, $T=60$ and $\lambda=10$ the Ledoit--Wolf estimator has
$\lambda_{\min}(\Sigh)=2.81\times10^{-4}$, so $\lambda\lambda_{\min}=2.81\times
10^{-3}$; the sample optimizer has $\norm{\phid^\star}_2=0.1796$; and the bound
equals $1.66$, $7.37$, $17.5$ and $25.4$ for
$\epsilon=10^{-3},10^{-4},10^{-5},10^{-6}$.

Three caveats follow from this distinction. First, the numbers just quoted are a
statement about the estimated problem and not a population guarantee; a
population breadth guarantee requires either knowledge of
$\lambda_{\min}(\Sigma)$ or a valid lower confidence bound for it, and we have
neither. Second, $\epsilon$ must be the gap of the objective that enters the
curvature constant, not the value guarantee of a surrogate problem. Third, with
the sample covariance the smallest eigenvalue vanishes when $T-1<N$, and then the
constant is zero and the bound is vacuous, so a breadth guarantee of this form is
available only in combination with a covariance regulariser whose spectral floor
can be quantified. That combination, and not the radius, is where the remaining
theoretical work lies.

\section{Routes we examined and set aside}

For transparency we record four routes that did not produce usable results.
(i) Replacing $\lambda_{\min}(\Sigh)$ by a curvature constant restricted to
feasible directions does not help, since the ratio is one on all 140 windows
(Table~\ref{tab:ashare}). (ii) The interior expansion of
Proposition~\ref{prop:affine} is exact but its domain is empty on our data: the
affine minimum-variance portfolio has negative weights, so the long-only
constraint is active at $\beta=0$ and there is no $\beta$ for which the
closed-form solution is feasible. (iii) Using the mean curvature
$\mathrm{tr}(\Sigma_{1^\perp})/(N-1)$ in place of the worst-case curvature would
improve the bound by a factor of order four, but it is not a lower bound on the
curvature in an arbitrary direction and requires distributional assumptions on
the error direction, which we were unable to establish. (iv) A cap on positions
does not make the norm penalty identically constant; the bound $\norm{\phid}_2\le
\sqrt{c}$ holds, but $(0.6,0.4,0)$ and $(0.6,0.2,0.2)$ both satisfy a cap of
$0.6$ with squared norms $0.52$ and $0.44$.

\section*{Data and code availability}

Every number in this paper is reproducible from two public artefacts. The
fourteen A-share return panels of Section~\ref{sec:evidence} are published, with
a manifest of per-file SHA-256 checksums, at
\url{https://huggingface.co/datasets/Changahou/GU2026_1}. The code that produces
the structural quantities, the audits, the rents and the figures, together with
the archived outputs from which the tables were generated and a log of the
implementation errors found while building them, is at
\url{https://github.com/ChenJiahaoJNU/GU2026-11}. What is redistributed is the
daily return matrix alone: no prices, identifiers, volumes or factor values, and
the upstream benchmark that produced the panels is cited in the dataset card.

\bibliographystyle{rQUF}
\bibliography{refs}

\end{document}